\documentclass[pra,twocolumn,aps,showpacs,superscriptaddress,floatfix,showpacs]{revtex4-1}

\usepackage{graphicx}
\usepackage{dcolumn}
\usepackage{bm}
\usepackage{latexsym}
\usepackage{amssymb}
\usepackage{amsmath}

\newcommand{\beq}{\begin{eqnarray*}}
\newcommand{\eeq}{\end{eqnarray*}}

\newcommand{\be}{\begin{eqnarray}}
\newcommand{\ee}{\end{eqnarray}}

\usepackage[usenames,dvipsnames]{color}

\usepackage[colorlinks,bookmarks=false,citecolor=blue,linkcolor=red,urlcolor=blue]{hyperref}

\begin{document}

\title{Scaling of the gap, fidelity susceptibility, and Bloch oscillations across the 
superfluid to Mott insulator transition in the one-dimensional Bose-Hubbard model}

\date{\today}

\author{Juan Carrasquilla}
\affiliation{Department of Physics, The Pennsylvania State University,
University Park, Pennsylvania 16802, USA}
\affiliation{Department of Physics, Georgetown University, Washington, DC 20057, USA}
\affiliation{Kavli Institute for Theoretical Physics, University of California,
Santa Barbara, California 93106, USA}
\author{Salvatore R. Manmana}
\affiliation{Institut f\"ur Theoretische Physik, Universit\"at G\"ottingen, Friedrich-Hund-Platz 1, 
D-37077 G\"ottingen, Germany}
\affiliation{JILA, (University of Colorado and NIST), and Department of Physics, 
University of Colorado, Boulder, Colorado 80309-0440, USA}
\affiliation{Kavli Institute for Theoretical Physics, University of California, 
Santa Barbara, California 93106, USA}
\author{Marcos Rigol}
\affiliation{Department of Physics, The Pennsylvania State University,
University Park, Pennsylvania 16802, USA}

\begin{abstract}
We investigate the interaction-induced superfluid-to-Mott insulator transition in the one-dimensional  
Bose-Hubbard model (BHM) for fillings $n=1$, $n=2$, and $n=3$ by studying the single-particle gap, the fidelity 
susceptibility, and the amplitude of Bloch oscillations via density-matrix renormalization-group 
methods. We apply a generic scaling procedure for the gap, which allows us to determine the critical points 
with very high accuracy.  We also study how the fidelity susceptibility behaves across the phase 
transition. Furthermore, we show that in the BHM, and in a system of spinless fermions, the amplitude of 
Bloch oscillations after a tilt of the lattice vanishes at the critical points. This indicates that 
Bloch oscillations can serve as a tool to detect the transition point in ongoing experiments 
with ultracold gases. 
\end{abstract}

\maketitle

\section{Introduction}
Ultracold atomic and molecular gases in optical lattices provide a unique playground for investigating 
quantum many-body phenomena \cite{bloch_dalibard_review_08,cazalilla_citro_review_11}. Since the seminal 
experiment by Greiner {\it et al.}~\cite{greiner_quantum_2002}, it has become common in such experiments 
to study quantum phase transitions in the presence of strong correlations. 
In particular, optical lattice realizations of the Bose-Hubbard model (BHM)
\begin{equation}
H = -J \sum\limits_{\langle ij\rangle} \left( b_i^\dagger b_{j}^{\phantom{\dagger}} + \textrm{H.c.} \right) + 
\frac{U}{2} \sum\limits_j n_j^{\phantom{\dagger}} \left(n_j^{\phantom{\dagger}} - 1\right),
\label{eq:BHM}
\end{equation} 
have been shown to undergo a transition from a superfluid to a Mott-insulator as the ratio of $U/J$ is 
increased in different dimensions \cite{greiner_quantum_2002,stoferle_moritz_04,spielman_phillips_07}.
In what follows, we set $J=1$ and $\hbar=1$, so that $U$ is measured in units of $J$ and time $t$ is
measured in units of $\hbar/J$. We also set the lattice spacing  $a=1$, thus length is measured in units of $a$. 

The one-dimensional (1D) BHM, the focus of this study, is of particular interest because of the dominant 
role played by quantum fluctuations. From the theoretical side, it is challenging to accurately determine 
the critical value $U_c$ at which the system at constant density undergoes a superfluid--Mott-insulator 
transition, something that, due to the lack of exact solutions for this model, is typically done utilizing 
computational approaches \cite{cazalilla_citro_review_11}. Here, the Berezinskii-Kosterlitz-Thouless (BKT) 
universality class of the transition makes calculations in finite systems susceptible to large finite-size 
effects. Insights from Luttinger liquid theory, combined with  density-matrix renormalization-group  (DMRG)
\cite{White:1992p2171,White:1993p2161,Schollwock:2005p2117,Schollwock:2011p2122} calculations of correlation 
functions and extrapolations to the thermodynamic limit, have provided some of the most accurate values of 
$U_c$ to date \cite{kuhner_white_00,ejima_dynamic_2011} (see Ref.~\cite{cazalilla_citro_review_11} for a 
review). Due to the large numerical effort needed, alternative and more accurate scaling approaches to 
calculate $U_c$, which do not rely on computing correlation functions, are highly desirable. 
From the experimental point of view, many of the quantities used in theoretical studies to determine $U_c$ 
are either difficult (e.g, the gap \cite{stoferle_moritz_04,clement_exploring_2009,roscilde_probing_2009,
PhysRevA.82.053609}) or not possible to measure accurately. The task is complicated even further by 
inhomogeneities induced by the unavoidable confining potentials present~\cite{batrouni_rousseau_02,wessel_04,
rigol_09}. Therefore, it is also highly desirable to find approaches to determine $U_c$ that could be more 
easily implemented in experiments. 

In this work, we address the two issues mentioned above, namely, how to accurately determine $U_c$ within 
computational approaches and in experimental studies. First, we apply a recently proposed scaling approach 
for the gap \cite{tapan_11} to obtain the critical point in the BHM with high accuracy and at fillings $n=1,\,2,\,3$. 
Second, we investigate the behavior of the fidelity susceptibility across these transitions. The fidelity 
susceptibility, a quantity motivated from the field of quantum information, has recently attracted much 
attention as a means of identifying the presence of quantum phase transitions even if the nature of the 
involved phases is not known \cite{zanardi_paunkovic_06,PhysRevLett.98.110601,PhysRevLett.99.095701,
PhysRevLett.99.100603,PhysRevB.76.180403,SirkerPRL2010,PhysRevB.78.115410,PhysRevB.77.245109,rigol_shastry_09,
varney_sun_11,SUN_paper,Michaud2012}. Finally, we discuss how to determine $U_c$ by studying the center-of-mass 
motion during Bloch oscillations, which occur after tilting the lattice \cite{PRL_tJVW,PhysRevA.84.033638}. 
This is something that can be easily implemented in ultracold gases experiments. 

All equilibrium calculations are done utilizing DMRG and, out-of-equilibrium, the Krylov variant of the 
adaptive time-dependent DMRG (t-DMRG) \cite{Schollwock:2005p2117}. For ground state calculations, we perform 
10 sweeps and keep up to $m=1000$ density-matrix eigenstates. In order to ensure the high accuracy needed 
for the considerations below, we truncate the local Hilbert space at $n+5$ bosons (where $n =1,\,2$ or $3$).  
The ground state energies obtained are converged in most cases with an absolute accuracy of $10^{-7}$ or 
better. The computation of the fidelity susceptibility is, however, more demanding, and we are restricted 
to smaller systems. In order to reach the necessary accuracy in the overlap of the two wave functions 
involved, we use $m \leq 4000$.  To study the out-of-equilibrium dynamics under a tilt, we truncate the
local Hilbert space at $n+4$ bosons and keep up to $m=2500$ states in the course of the time evolution 
while using a time step of $\Delta t = 0.01$.    

\section{Scaling analysis of the gap}

The phase transition from a Mott insulator to a superfluid in the 1D BHM at commensurate fillings is 
known to be of BKT type \cite{PhysRevB.40.546}. Hence, it is accompanied by the exponential closing of the 
single-particle gap $E_g\sim \exp(-b/\sqrt{U-U_c})$ ($b$ is a parameter which is independent of $U$). As a 
consequence of its exponential behavior, a direct study of the transition by computing the single-particle 
gap for finite systems is plagued by finite-size effects.  This problem can be overcome by a scaling analysis 
of the gap, for which we follow the approach in Ref.~\cite{tapan_11}, briefly described below. 

The method is based on the following ansatz for the scaling of the gap in the vicinity of the phase transition,
\begin{equation}
L E_g\left(L\right) \times \left(1+\frac{1}{2\ln{L}+C} \right)= F\left( \frac{\xi}{L} \right),
\label{eq:scaling}
\end{equation}
where $F$ is a scaling function, $C$ is an unknown constant to be determined, and $L$ is the system size.
We emphasize two aspects of this scaling ansatz: First, it contains the logarithmic corrections that 
are typical for $E_g\left(L\right)$ at the BKT transition \cite{0305-4470-20-2-010,PhysRevB46_7422}. 
Second, it resembles the relation for the resistance (which also vanishes exponentially) in the 
charge-unbinding transition of the two-dimensional classical Coulomb gas, which is also of BKT type 
\cite{PhysRevB.51.6163}. At the critical point, and in its vicinity within the superfluid region, one expects 
the values of $F(\xi/L)$ to be system-size independent because of the divergence of the correlation length.
Hence, the data for the rescaled gap $ E_g^{*}\left(L\right)=L E_g\left(L\right) \left[1+1/\left(2\ln{L}+
C\right) \right]$ for different system sizes $L$ will be independent of $L$ in this region. Furthermore, 
the curves $E_g^{*}\left(L\right)$ vs $\xi/L$ for several values of $L$ and $U$ should collapse onto a 
unique curve representing $F$. Equivalently, one can reformulate the relation in Eq.~\eqref{eq:scaling} 
by taking the logarithm of the argument of $F$ and considering a different function $f$ with argument 
$x_L=\ln L -\ln \xi$. 

We determine the critical point by adjusting the parameters $U_c$, $b$, and $C$. In the procedure, we 
look for the best collapse of the curves $E_g^{*}\left(L\right)$ vs $x_L$ for different values of $U$ and 
$L$.  This is done by representing the function $f$ with a selected high-degree polynomial (eighth degree in 
our case) such that the results are independent of the degree. Such polynomial is fit on a dense grid of values 
of $U_c$, $b$, and $C$, to the calculated values of $E_g^{*}\left(L\right)$ and $x_L$.  The quality of the fit 
is assessed by computing the sum of squared residuals ($S$), which defines the function $S(U_c,b,C)$. $U_c$ 
is then obtained from the set of parameters $U_c$, $b$, and $C$ which minimizes $S(U_c,b,C)$. The accuracy of 
this method was tested by locating the critical interaction strength in a model of spinless fermions with 
nearest-neighbor interaction. Such a model exhibits a BKT transition for which the critical interaction strength 
is known analytically \cite{tapan_11}. The value of the critical interaction strength obtained utilizing the
scaling analysis of the gap described before deviates only $1\%$ from the exact value~\cite{tapan_11}, and thus 
we are confident that a similar or better accuracy should be attained for the BKT transition in the BHM. 

\begin{figure}[!t]
\includegraphics[width=0.59\textwidth,angle=-90]{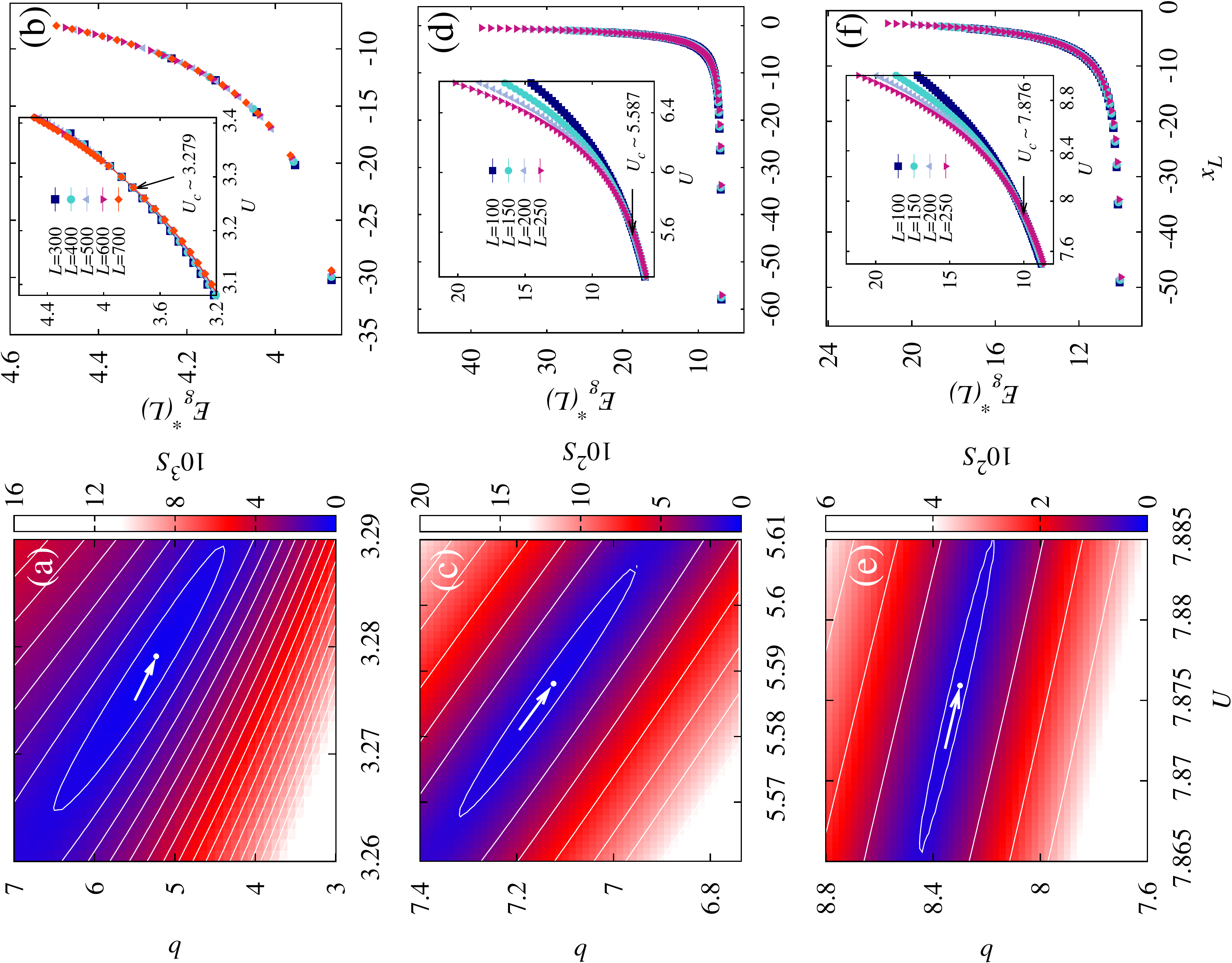}
\caption{(Color online) (a) Contour plot of the sum of squared residuals $S(b,U_c)$ for $n=1$. The white 
arrow signals the location of the minimum value of $S$. The white lines are equally spaced contour lines 
where $S$ is constant. (b) Best collapse of the data for $E_g^{*}\left(L\right)$ vs $x_L$ 
corresponding to $U_c=3.279$, $b=5.2$, and  $C\to\infty$. The inset shows the rescaled gap vs. $U$. 
A similar analysis for $n=2$ [$n=3$] is presented in panels (c) and (d) [(e) and (f)]. $U$ and $E_g^{*}$ are
presented in units of $J$, whereas  $b$ and $S$ are shown in units of $J^{1/2}$ and $J^2$, respectively.}
\label{fig:gap_ssr}
\end{figure}
 
We have applied this procedure to integer filled chains with $n = 1,\, 2,$ and $3$. We find that, in these 
three cases, the minimum of $S(U_c,b,C)$ is obtained for arbitrarily large values of $C$. This means that 
logarithmic corrections to the scaling of the gap, in the form \eqref{eq:scaling}, do not play a role in 
the determination of the critical point. This is to be contrasted with the $t$-$V$-$V'$ model in 
Ref.~\cite{tapan_11}, where $C$ was found to be finite in all transitions analyzed. In 
Fig.~\ref{fig:gap_ssr}(a), we present a density plot corresponding to $S(U_c,b,\infty)$ for $n=1$, which 
exhibits a clear minimum at $U_c=3.279\pm0.001$, $b=5.2\pm0.1$. The error bars are estimated by repeating the
minimization procedure adding and subtracting to the gap the error of the energy (overestimated to be 
$\sigma_{E_g}=10^{-6}$). Furthermore, the sensitivity of the results to the selection of the interval of
values of $U$ used in the fit is also included in the error bars such that our results are independent of that 
choice. Corresponding to the set of parameters that minimizes $S(U,b,C)$, in Fig.~\ref{fig:gap_ssr}(b), we plot 
$E_g^{*}\left(L\right)$ vs $x_L$. The data are clearly seen to collapse to a single curve representing the 
function $f$. In the inset, the curves for the rescaled gap corresponding to different system sizes are seen 
to merge around the critical value $U_c$. 

Previous calculations have obtained $U_c$ through widely different techniques, some of which we mention 
below. An early quantum Monte Carlo (QMC) study found $U_c=4.7\pm0.2$ using the closing of the gap 
at the critical point \cite{PhysRevLett.65.1765,PhysRevB.46.9051}, while later QMC simulations yielded a 
smaller value $U_c=3.33\pm0.06$~\cite{jetp2}. An approximated calculation using the Bethe ansatz suggested 
that $U_c=3.460$~\cite{PhysRevB.44.9772}, and strong coupling expansion calculations predicted 
$U_c=3.8\pm0.1$~\cite{PhysRevB.59.12184}. Exact diagonalization studies led to $U_c=3.64\pm0.07$ 
\cite{jetp1}, while combining exact diagonalization with renormalization group insights, a value of 
$U_c=3.28\pm0.02$ was reported in Ref.~\onlinecite{PhysRevB.53.11776}. 
Also, using extrapolated measurements of the fidelity susceptibility 
extracted from exact diagonalization of small clusters, $U_c=3.89\pm0.02$ was found in 
Ref.~\cite{PhysRevLett.98.110601}. In Ref.~\cite{PhysRevLett.76.2937}, one of the first DMRG approaches 
to tackle this problem using extrapolations of the gap, a value of $U_c=3.36$ was determined. 
Later DMRG studies, based on accurate extrapolations of the decay of correlation functions, reported 
$U_c=3.6\pm0.1$~\cite{PhysRevB.58.R14741} and $U_c=3.3\pm0.1$~\cite{kuhner_white_00}, and, more recently, 
$U_c=3.361\pm0.006$ \cite{Zakrzewski_Delande_2008} and $U_c=3.27\pm0.01$ \cite{ejima_dynamic_2011}. 
Computing the Luttinger parameter using bipartite fluctuations, Ref.~\cite{RachelPRL2012} reported 
$U_c = 3.345\pm0.003$. Finite-size scaling analyses of the von Neumann entanglement entropy suggested that
$U_c$ lays between $U=3.3$ and $U=3.4$ ~\cite{1742-5468-2008-05-P05018} and 
$U_c=3.27\pm0.03$~\cite{PhysRevA.85.053644}, while computations of the von Neumann entropy directly in the 
thermodynamic limit (using the infinite time-evolving block decimation algorithm) produced a critical value 
$U_c=3.3\pm0.1$ \cite{PhysRevA.86.023631}. Our result for $U_c$ is therefore in good agreement with the 
lowest values reported in the most recent studies that use widely diverse quantities to characterize 
this transition.

We have also computed the critical values for other commensurate fillings, and found $U_c=5.587\pm0.001$ 
for $n=2$ [Figs.~\ref{fig:gap_ssr}(c) and \ref{fig:gap_ssr}(d)], and $U_c=7.876\pm0.002$ for $n=3$ 
[Figs.~\ref{fig:gap_ssr}(e) and \ref{fig:gap_ssr}(f)]. Note that our value of $U_c$ for 
$n=2$ is in excellent agreement with the large-scale DMRG study in Ref.~\cite{ejima_dynamic_2011}, 
further supporting that the scaling of the gap utilized here is capable of providing very accurate results 
at a lower computational cost. In what follows, we use our results for $U_c$ to benchmark alternative 
approaches for locating the transition point. 
 
\section{Fidelity Susceptibility}

The fidelity susceptibility (FS) $\chi$ for the ground state of the system $|\psi_0\rangle$ is defined as 
\begin{equation}
\chi(U) =  \frac{2 \left[ 1 -  \left| \langle \psi_0(U) | \psi_0(U+dU) \rangle \right| \right]}{L \, dU^2},
\label{eq:chi}
\end{equation}
and is also known as the fidelity metric. For generic second-order phase transitions, $\chi$ is expected 
to diverge in the thermodynamic limit (TL) \cite{zanardi_paunkovic_06,PhysRevLett.99.095701,
PhysRevLett.99.100603,PhysRevB.76.180403,PhysRevB.78.115410}, and it has been found to exhibit clear 
signatures of such transitions already for rather small system sizes, where a maximum of $\chi$ was 
seen near the transition point \cite{PhysRevB.77.245109,rigol_shastry_09,varney_sun_11,Michaud2012}. 

\begin{figure}[!b]
\includegraphics[width=0.43\textwidth]{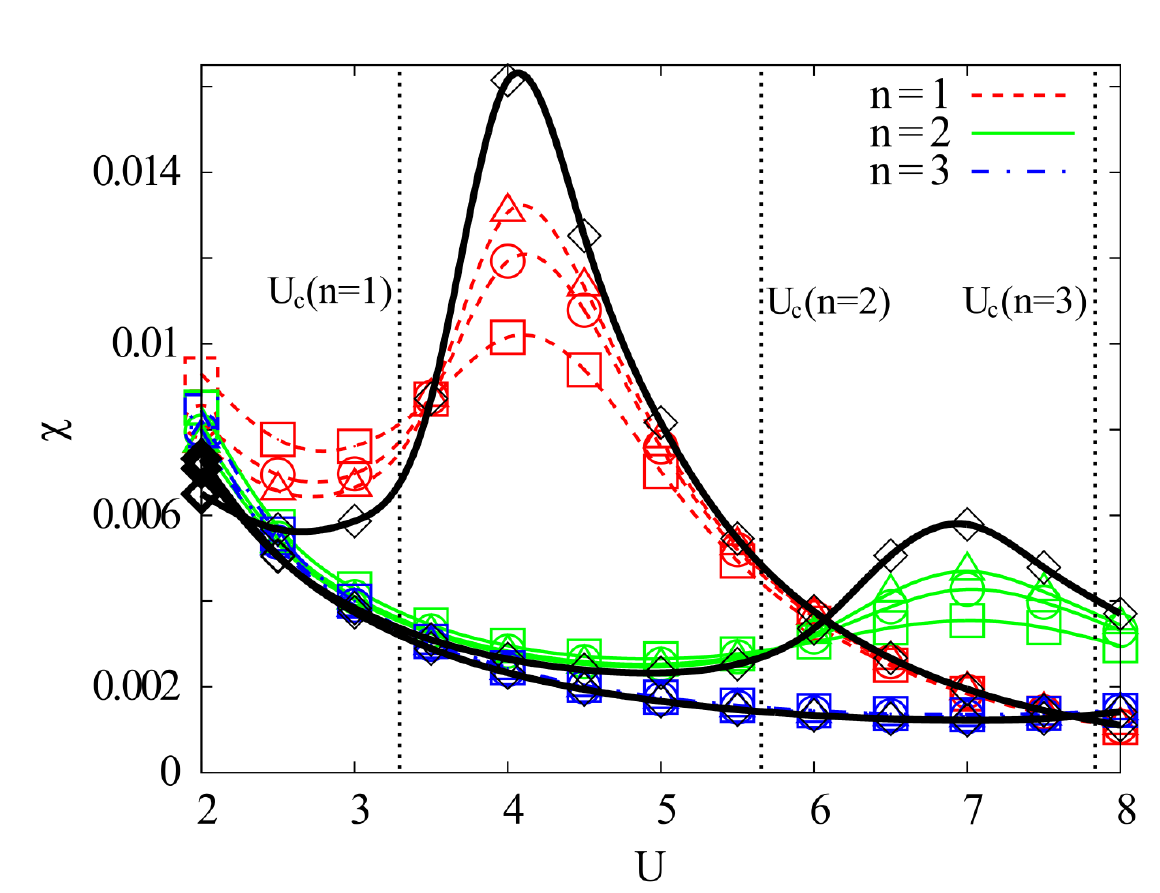}
\caption{(Color online) Fidelity susceptibility for different system sizes at integer filling $n=1$ 
(red symbols and dashed lines), $n=2$ (green symbols and solid lines) and $n=3$ (blue symbols and 
dot-dashed lines). The plot shows data for $L=40$ (square), $L=80$ (circle) and $L=120$ 
(triangle), the lines are spline interpolations and serve as a guide to the eye. The thick solid black lines 
(diamonds) are the result of a finite-size extrapolation using a quadratic fit. The vertical dotted lines 
indicate the position of the quantum critical points obtained using the scaling analysis of the gap 
described in the text. $U$ and $\chi$ are presented in units of $J$ and $J^{-2}$, respectively.} 
\label{fig:fidsusc}
\end{figure}

In Fig.~\ref{fig:fidsusc}, we show the fidelity susceptibility for the BHM at fillings $n=1,\,2,\,3$, 
for systems with $L=40,\,80,\,120$, and for on-site interactions up to $U=8$. For $n=1$ and $n=2$, $\chi$ 
exhibits clear maxima for values of $U$ greater than $U_c$ computed from the scaling of the gap. 
Consistent with the results in Ref.~\cite{PhysRevLett.98.110601}, the positions of the maxima are seen to
move toward weaker interactions, and their height to increase, with increasing system size.
For $n=3$, the maxima are expected to be beyond the values of $U$ studied here. Hence, indications for the 
existence of a phase transition are obtained already for small systems. Interestingly, and also of relevance 
to the $\chi$'s calculated here, recent works have proposed that a minimum of the FS may signal the quantum 
critical point \cite{PhysRevE.76.022101,SUN_paper}. This was argued to be possible because, depending on 
the scaling dimensions of the system, the FS can be finite at a critical point \cite{PhysRevLett.99.095701}. 
In Fig.~\ref{fig:fidsusc}, one can indeed see that minima of $\chi$ also occur close to the critical point. 

It is also apparent in our results in Fig.~\ref{fig:fidsusc} that, between the maxima and the minima, there 
is a point at which all values of $\chi$ seem to be independent of $L$ for the system sizes treated. A similar 
scenario was observed in the XXZ chain \cite{SirkerPRL2010} and for SU($N$) Hubbard chains \cite{SUN_paper}. 
As seen in Fig.~\ref{fig:fidsusc}, for $n=1$ and $n=2$, the ``crossing'' of the FS curves for different 
system sizes occurs at values of $U$ greater than $U_c$ computed from the scaling of the gap. We applied 
different extrapolation schemes for $\chi$ and did not obtain results consistent with those from 
the scaling of the gap: For example, in Fig.~\ref{fig:fidsusc} we show the outcome of the simplest 
approach in which the extrapolation to the thermodynamic limit is attained using a second-order 
polynomial. Neither the position of the maximum, nor the one of the minimum or of the crossing point, 
is in agreement with the values of $U_c$ obtained from the scaling of the gap.  

As pointed out in previous studies (see, e.g., Ref.~\cite{PhysRevB.78.115410} for an analysis of the 1D Fermi 
Hubbard model), the divergence of $\chi$ can be extremely slow, and very large system sizes (as well as a 
more elaborate finite-size-scaling ansatz) may be required to resolve the critical point. This is further 
supported by the results in Ref.~\cite{SirkerPRL2010}, in which a field theoretical analysis of $\chi$ at the 
BKT transition in the XXZ chain unveiled a very slow divergence. However, the numerical findings for the BHM 
here and the XXZ chain in Ref.~\cite{SirkerPRL2010} differ in two important aspects: as opposed to the behavior of the XXZ chain, in this work we have found that logarithmic corrections are negligible in the finite-size scaling of 
the gap of the BHM (for the system sizes analyzed). Also, the crossing point of the FS in the XXZ chain occurs 
for a value of the interaction strength that is smaller than the critical one, which is the opposite of what 
we find here for the BHM. Hence, the numerical data at hand make it difficult to determine $U_c$ utilizing 
the FS; further studies are needed to fully understand the behavior of this quantity in the BHM, and in 
particular, its contrast to the one observed for the XXZ chain.

\section{Center-of-mass motion}

In order to determine the critical point in experiments with ultracold bosons in optical lattices, we propose 
to follow a recent proposal that uses Bloch oscillations \cite{PRL_tJVW}. The idea is to apply an external field 
\begin{equation}
V_{\rm tilt} = - \Omega \sum\limits_j^{L} \, j\, n_j^{\phantom{\dagger}},
\end{equation}
at time $t\geq 0$, to a system that is initially in its ground state ($\Omega=0$ for $t<0$). Such a set up can 
be realized in optical lattice experiments by, e.g., tilting the lattice. One can then study the center-of-mass 
motion (COM)
\begin{equation}
x_{\rm CM}(t) = \frac{1}{N} \sum\limits_j^{L} \, j \, \langle n_j \rangle_t, 
\end{equation} 
where $N$ is the total number of particles, at times $t\geq 0$. In previous studies, in a variant of the 
$t$-$J$ model at low filling \cite{PRL_tJVW} and in an effective Ising model in a transverse field 
\cite{PhysRevA.84.033638}, it was reported that the amplitude of the COM exhibited signatures of the quantum 
phase transition at the critical point. In the $t$-$J$ like model \cite{PRL_tJVW}, because of the formation 
of pairs, the transition from a metallic to a gapped superconducting phase was visible by both a rapid 
decrease of the amplitude and by a doubling of the frequency of the Bloch oscillations. In the Ising like 
model \cite{PhysRevA.84.033638}, the amplitude was found to be maximal at the transition point. From the 
experimental point of view, Bloch oscillations have been, e.g., used to investigate Dirac points on hexagonal 
optical lattices \cite{Tarruell2012}, as well as to study low-frequency breathing modes in elongated Fermi 
gases~\cite{PhysRevLett.103.170402}. Here, we investigate what happens in the BHM and, at the same time, 
analyze the simpler (integrable) case of spinless fermions with nearest-neighbor interaction $V$, 
\begin{equation}
H = -J \sum\limits_j \left( c_{j+1}^\dagger c_j^{\phantom{\dagger}} + \textrm{H.c.} \right) + 
V \sum\limits_j n_j^{\phantom{\dagger}} n_{j+1}^{\phantom{\dagger}}.
\label{eq:tVmodel}
\end{equation}
As mentioned before, this model is exactly solvable and, at half filling, exhibits a BKT transition from 
a Luttinger liquid (LL) to a charge density wave (CDW) insulator at $V/J=2$ 
\cite{cazalilla_citro_review_11,tapan_11}. In what follows, we set $J=1$ and $\hbar=1$, so that $V$ is given 
in units of $J$ and time $t$ in units of $\hbar/J$. 
 
\begin{figure}[!t]
\includegraphics[width=0.45\textwidth]{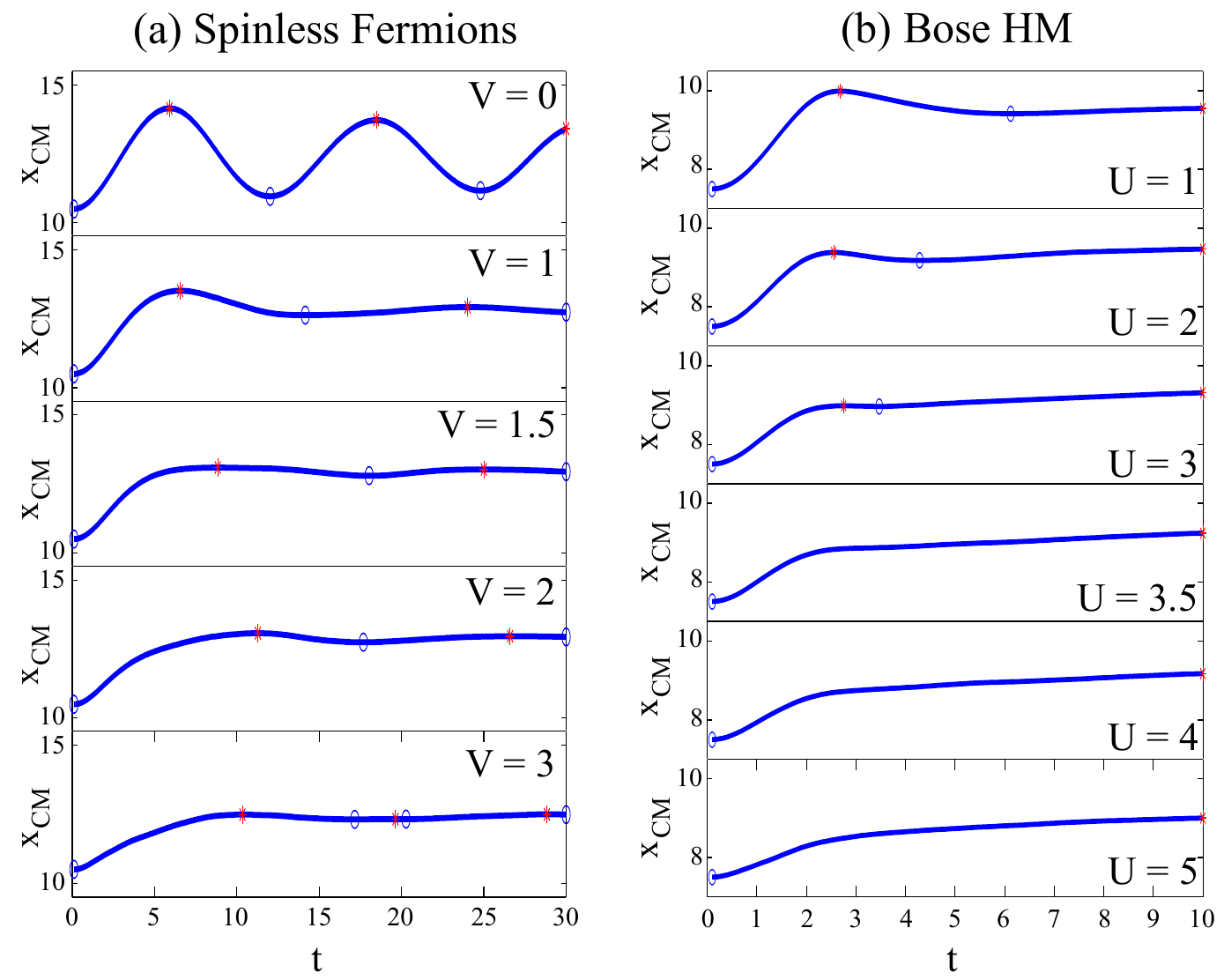}
\caption{(Color online)  COM for (a) spinless fermions ($L = 20$, $\Omega= 1$) after 
tilting the lattice for the indicated values of $V = 0,\ldots,3$. (b) The BHM ($L = 14$, $\Omega = 1$) 
after tilting the lattice for the indicated values of $U = 1,\ldots,5$. The red stars and blue circles 
denote maxima and minima of the oscillations, respectively. The center of mass is presented in units of the 
lattice spacing $a$, while time $t$ is measured in units of $\hbar/J$.}
\label{fig:oscillations_tV}
\end{figure}

In Fig.~\ref{fig:oscillations_tV}(a), we display the COM of a half-filled chain of spinless fermions with 
$L=20$ and different values of $V$ on both sides of the LL to CDW transition. It is apparent that the amplitude 
of the oscillations decreases and the damping rate increases when increasing $V$ and, deep in the CDW phase, 
no oscillations can be resolved. This is reminiscent of the behavior of a harmonic oscillator which moves 
freely ($V = 0$), damped ($0 < V \lesssim 2$) and overdamped ($V \gtrsim 2$). In the LL phase, mass transport 
is ballistic and, therefore, it is possible for spinless fermions to freely flow upon the introduction of a 
small tilt of the lattice, which gives rise to COM oscillations. On the other hand, in the CDW phase the 
system is gapped and transport under a small tilt is suppressed, which precludes COM oscillations. A 
qualitatively similar behavior is observed in the COM of the BHM at filling $n=1$ and for $1 \leq U \leq 5$ 
[Fig.~\ref{fig:oscillations_tV}(b)]. There, finite values of $U\lesssim 3.5$ lead to damped oscillations, 
and only the first oscillation can be resolved on the time scale of our simulations. For $U \geq 3.5$, 
overdamped behavior sets in and no oscillations can be identified.

\begin{figure}[!t]
\includegraphics[width=0.35\textwidth,angle=-90]{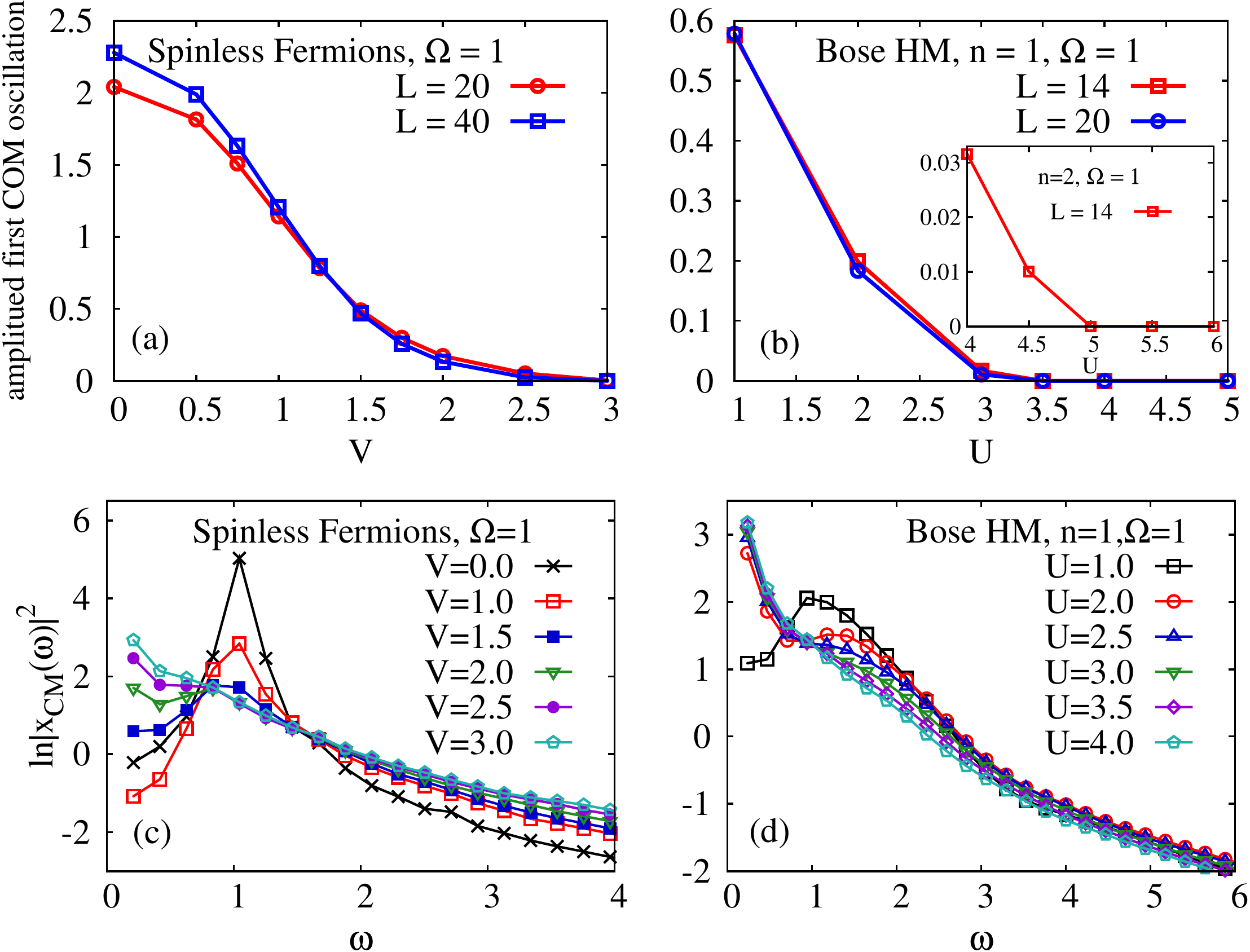}
\caption{(Color online) Amplitude of the first oscillation of the COM after tilting the lattice. 
(a) Spinless fermions at half-filling ($L = 20$ and $L=40$) and $\Omega = 1$. (b) BHM at filling $n=1$ 
($L = 14$ and $L=20$) and $\Omega = 1$. The inset shows results for $L=14$ at filling $n=2$. (c) The 
logarithm of the amplitude square of the Fourier transform of the COM oscillations for 
spinless fermions at half-filling with $L = 20$ and $\Omega = 1$. (d) Same as in (c) 
but for the BHM at $n=1$ with $L = 14$ and  $\Omega = 1$. The amplitudes of COM oscillations are 
measured in units of $a$, $V$ and $U$ in units of $J$, and $\omega$ in units of $J/\hbar$}
\label{fig:BO_amplitudes}
\end{figure}

To gain a better understanding of the evolution of the Bloch oscillations as interactions are increased, 
in Fig.~\ref{fig:BO_amplitudes}, we display the amplitude (defined as the difference between the first 
maximum and the first minimum), for spinless fermions vs $V$ [Fig.~\ref{fig:BO_amplitudes}(a)] and for the 
BHM vs $U$ [Fig.~\ref{fig:BO_amplitudes}(b)]. For spinless fermions, it can be seen that the amplitude of the 
oscillations at the critical point and above ($V\geq2$) is very small and decreases with increasing system 
size. The results for the BHM are qualitatively similar. The region of $U$ at which the amplitude of the 
Bloch oscillations is seen to vanish, $3.0<U\leq 3.5$, contains the value obtained from the scaling analysis 
of the gap $U_c \approx 3.279$. 

This behavior of the Bloch oscillations is also reflected in the Fourier transform (FT) of the time 
evolution of the COM, which we present in Fig.~\ref{fig:BO_amplitudes}(c) for spinless fermions and in
Fig.~\ref{fig:BO_amplitudes}(d) for the BHM. In both cases, for weak interactions, there is a well-defined 
peak around $\omega\sim 1$, which reflects the oscillations observed in Fig.~\ref{fig:oscillations_tV}. 
As the interaction strength is increased, the height of that peak slowly decreases and its position (slightly) 
changes. This is accompanied by an increase in the weight of the zero-frequency mode. For both systems, 
as the interaction is increased past the critical value, it is no longer possible to resolve the finite 
frequency peak. This is another indication that the COM oscillations are suppressed for $U\geq U_c$. Therefore, 
for both systems, the BKT transition leads to comparable behavior, and the study of Bloch oscillations in 
experiments can provide a good estimate of $U_c$. 

We also studied the COM for $n=2$ and $n=3$. Since computations become increasingly demanding 
with increasing filling, only smaller lattice sizes could be studied in those cases. 
In addition, the numerical values of the amplitude become significantly smaller.
We therefore find that theoretical estimates for $U_c$ from the behavior of the Bloch oscillations 
become less accurate with increasing $n$. In the inset in Fig.~\ref{fig:BO_amplitudes}(b), we show 
results for $n=2$ and $L=14$, where one can see that the amplitude of the Bloch oscillations vanishes 
for $4.5<U\leq5$, while the theoretical prediction is $U_c=5.59$. It would be interesting to
study Bloch oscillations in larger lattice sizes in experiments and see if the worsening of the 
predictions is due to finite-size effects or due to a worsening of this 
approach with increasing filling.\\

\section{Conclusions and Outlook}
We followed three approaches to study quantum critical behavior in the one-dimensional BHM at integer 
fillings. By means of a scaling analysis of the gap, we obtained accurate values of $U_c$ for the superfluid 
to Mott insulator transition at fillings $n=1,\,2,\,3$. The fidelity susceptibility was shown to exhibit 
signatures of the phase transitions for finite systems, but the results for this quantity did not allow us 
to improve on the values of $U_c$ obtained from the scaling of the gap. Finally, we showed that the study of Bloch 
oscillations in experiments can help locating the critical values for the superfluid-to-Mott-insulator 
transition. The latter approach could potentially be used also in experiments in higher dimensions.   

\begin{acknowledgments}
This work was supported by the Office of Naval Research and by the National Science Foundation under 
Grants No.~PHY11-25915 and PIF-0904017, and in part under Grant No.~PHY05-25915. We thank 
J. Sirker, F. Gebhard, L. Campos Venuti, L. Tarruell, K. Hazzard, G. Chen, A. M. Rey, S. Ejima, and N. Bray-Ali for discussions, 
and acknowledge the use of computational resources at the Janus supercomputer (CU Boulder).
\end{acknowledgments}

\bibliographystyle{prsty}


\end{document}